\documentclass[aps,twocolumn,showpacs,showkeys]{revtex4}
\usepackage{amsmath}
\usepackage[dvips]{graphicx}
\usepackage{amsfonts}
\usepackage{amssymb}

\begin{document}

\title{A detailed analysis of dipolar interactions and analytical
approximations in arrays of magnetic nanowires}
\author{D. Laroze}
\affiliation{Instituto de F\'{\i}sica, Pontificia Universidad Cat\'{o}lica de Valpara%
\'{\i}so, Av. Brasil 2950, Valpara\'{\i}so, Chile.}
\author{J. Escrig, P. Landeros and D. Altbir}
\affiliation{Departamento de F\'{\i}sica, Universidad de Santiago de Chile, USACH, Av.
Ecuador 3493, Santiago, Chile.}
\author{M. V\'{a}zquez}
\affiliation{Instituto de Ciencia de Materiales-CSIC, Campus de Cantoblanco, 28049
Madrid, Spain.}
\author{P. Vargas}
\affiliation{Departamento de F\'{\i}sica, Universidad T\'{e}cnica Federico Santa Mar\'{\i}%
a, Casilla 110-V, Valpara\'{\i}so, Chile.}
\keywords{nanowires, nanomagnetism, magnetostatic interaction, wire arrays,
dipolar interaction.}
\pacs{75.75.+a,75.10.-b}

\begin{abstract}
The investigation of the role of interactions in magnetic wire arrays is
complex and often subject to strong simplifications. In this paper we
obtained analytical expressions for the magnetostatic interactions between
wires and investigate the range of validity of dipole-dipole, first order
and second order approximations. We also analyze the extension of the
interwire magnetostatic interactions in a sample and found that the number
of wires required to reach energy convergence in the array strongly depends
on the relative magnetic orientation between the wires.
\end{abstract}

\maketitle

\section{Introduction}

During the last decade, regular arrays of magnetic nanoparticles have been
deeply investigated. Besides the basic scientific interest in the magnetic
properties of these systems, there is evidence that they might be used in
the production of new magnetic devices. \cite{Prinz, Cowburn} Different
geometries have been considered, including dots,\ rings, tubes and wires.
Recent studies on such structures have been carried out with the aim of
determining the stable magnetized state as a function of the geometry of the
particles. \cite{Jubert, Porrati, Landeros, Escrig2} In particular, the
study of highly ordered arrays of magnetic wires with diameters typically in
the range of tens to hundred nanometers is a topic of growing interest. \cite%
{Nielsch3, Nielsch4, Vazquez, Vazquez2} This is a consequence of the
development of experimental techniques that lead to fabricate in a
controllable and ordered way such arrays. \cite{Martin, Ross2} The high
ordering in the array, together with the magnetic nature of nanowires, give
rise to outstanding cooperative properties of fundamental and technological
interest. \cite{Skomski}

Bistable nanowires are characterized by squared-shaped hysteresis loops
defined by the abrupt reversal of the magnetization between two stable
remanent states. \cite{Varga, Sampaio} In such systems, effects of
interparticle interactions are in general complicated by the fact that the
dipolar fields depend upon the magnetization state of each element, which in
turn depends upon the fields due to adjacent elements. Therefore, the
modelling of interacting arrays of nanowires is often subject to strong
simplifications like, for example, modelling the wire using a
one-dimensional modified classical Ising model. \cite{Sampaio, Knobel} Zhan
\textit{et al. }\cite{Zhan} used the dipole approximation including
additionally a length correction. J. Vel\'{a}zquez and M. V\'{a}zquez \cite%
{Velazquez1, Velazquez2} considered each microwire as a dipole, in a way
that the axial field generated by a microwire is proportional to the
magnetization of the microwire. Nevertheless, this model is merely
phenomenological since the comparison of experimental results with a
strictly dipolar model shows that the interaction in the actual case is more
intense. They have also calculated the dipolar field created by a cylinder
and expanded the field in multipolar terms, \cite{Velazquez} showing that
the non-dipolar contributions of the field are non negligible for distances
considered in experiments. In spite of the extended use of the dipole-dipole
approximation, a detailed calculation of the validity of approximations on
the dipolar energy has not been presented yet. Also micromagnetic
calculations \cite{Hertel, Clime} and Monte Carlo simulations \cite{Bahiana}
have been developed. However, these two methods permit to investigate arrays
with just a few wires. In this paper we investigate the validity of the
dipole-dipole approximation and include additional terms that lead us to
consider large arrays and the shape anisotropy of each wire.

\section{Continuous magnetization model}

Geometrically, nanowires are characterized by their radius, $R$, and length,
$L$. The description of an array of $N$ wires based on the investigation of
the behavior of individual magnetic moments becomes numerically prohibitive.
In order to circumvent this problem we use a continuous approach and adopt a
simplified description in which the discrete distribution of magnetic
moments in each wire is replaced with a continuous one, defined by a
function $\mathbf{M}(\mathbf{r})$ such that $\mathbf{M}(\mathbf{r})\delta V$
gives the total magnetic moment within the element of volume $\delta V$
centered at $\mathbf{r}$. We recall that $E_{tot}$ is generally given by the
sum of three terms corresponding to the magnetostatic, $E_{dip}$, the
exchange, $E_{ex}$, and the anisotropy contributions. Here we are interested
in soft or polycrystalline magnetic materials, in which case the anisotropy
is usually disregarded. \cite{Nielsch4}

The total magnetization can be written as $\mathbf{M}(\mathbf{r}%
)=\sum_{i=1}^{N}\mathbf{M}_{i}(\mathbf{r})$, where $\mathbf{M}_{i}(\mathbf{r}%
)$ is the magnetization of the $i$-th nanowire. In this case, the
magnetostatic potential $U(\mathbf{r)}$ splits up into $N$ components, $%
U_{i}(\mathbf{r)}$, associated with the magnetization of each individual
nanowire. Then, the total dipolar energy can be written as $%
E_{dip}=\sum_{i=1}^{N}E_{dip}(i)+\sum_{i=1}^{N-1}%
\sum_{j=i+1}^{N}E_{int}(i,j) $, where
\begin{equation}
E_{dip}(i)=\frac{\mu _{0}}{2}\int \mathbf{M}_{i}(\mathbf{r})\nabla U_{i}(%
\mathbf{r})\,dv  \label{1}
\end{equation}%
\noindent is the dipolar contribution to the self-energy of nanowire $i$-$th
$, and
\begin{equation}
E_{int}\left( i,j\right) =\mu _{0}\int \mathbf{M}_{i}(\mathbf{r})\nabla
U_{j}(\mathbf{r})\,dv  \label{2}
\end{equation}%
\noindent is the dipolar interaction between them. In the dipolar
contribution to the self-energy an additive term independent of the
configuration has been left out. \cite{Aharoni}

In this work we investigate bi-stable nanowires in which case, \cite{Aharoni}
$E_{ex}=\sum_{i=1}^{N}E_{ex}(i)=0.$ On the basis of this result, the total
energy of the array can be written as
\begin{equation}
E_{tot}=\sum_{i=1}^{N}E_{self}(i)+\sum_{i=1}^{N-1}%
\sum_{j=i+1}^{N}E_{int}(i,j),  \label{4}
\end{equation}%
\noindent where $E_{self}(i)=E_{dip}(i)$ is the dipolar self-energy of each
wire, and $E_{int}(i,j)$ is the dipolar interaction energy between wires $i$-%
$th$ and $j$-$th$.

\subsection{Total energy calculation}

We now proceed to the calculation of the energy terms in the expression for $%
E_{tot}$. Results will be given in units of $\mu _{0}M_{0}^{2}V$, i.e. $%
\tilde{E}=E/\mu _{0}M_{0}^{2}V$, where $V=\pi R^{2}L$ is the volume of the
nanowire and $M_{0}$ is the saturation magnetization.

In order to evaluate the total energy, it is necessary to specify the
functional form of the magnetization for each nanowire. We consider wires
with an axial magnetization defined by $\mathbf{M}_{i}(\mathbf{r}%
)=M_{0}\sigma _{i}\mathbf{\hat{z}}$, where $\mathbf{\hat{z}}$ is the unit
vector parallel to the axis of the nanowire and $\sigma _{i}$ takes the
values $\pm 1$, allowing the wire $i$ to point up ($\sigma _{i}=+1$) or down
($\sigma _{i}=-1$) along$\ \mathbf{\hat{z}}$.

\subsubsection{Self energy of a nanowire}

The reduced dipolar self-energy has been calculated by Landeros \textit{et
al.} \cite{Landeros} and takes the form%
\begin{equation}
\tilde{E}_{self}(i)=\frac{1}{2}\left( 1+\frac{8R}{3\pi L}-F_{21}\left[ -%
\frac{4R^{2}}{L^{2}}\right] \right) \ ,  \label{5}
\end{equation}%
\noindent where $F_{21}[x]=F_{21}[-1/2,1/2,2,x]$ is a hypergeometric
function. Note that in Eq. (\ref{5}) the energy of each wire depends only on
the ratio $R/L$.

\subsubsection{Interwire magnetostatic coupling}

The interaction between two nanowires is obtained using the magnetostatic
field experienced by one of the wires due to the other. Details of these
calculations are included in Appendix A, giving%
\begin{equation}
\tilde{E}_{int}(i,j)=2\sigma _{i}\sigma _{j}\int\limits_{0}^{\infty }\frac{dq%
}{q^{2}}J_{0}\left( \frac{qS_{ij}}{L}\right) J_{1}^{2}\left( \frac{qR}{L}%
\right) (1-e^{-q})\ ,  \label{6}
\end{equation}%
\noindent where $J_{p}$ is a Bessel function of first kind and $p$ order and
$S_{ij}$ is the center-to-center distance between the magnetic nanowires $i$
and $j$. The previous equation allows us to write the interaction energy of
two wires as $\tilde{E}_{int}(i,j)=\sigma _{i}\sigma _{j}\tilde{E}%
_{int}(S_{ij})=\pm \tilde{E}_{int}(S_{ij})$, where the sign $+$ $(-)$ \
corresponds to $\sigma _{i}=\sigma _{j}$ $(\sigma _{i}\neq \sigma _{j})$,
respectively.

\subsection{Results}

\subsubsection{Two wires system}

The general expression giving the interaction between wires with axial
magnetization is giving by Eq. (\ref{6}). This expression has to be solved
numerically. However, wires that motivate this work \cite{Nielsch3,
Nielsch4, Vazquez, Vazquez2} satisfy $L/R\gg 1$, leading us to expand $J_{1}$
as%
\begin{equation}
J_{1}(x)=\frac{x}{2}-\frac{x^{3}}{16}+\sum_{k=2}^{\infty }\frac{%
(-1)^{k}(x/2)^{1+2k}}{k!\Gamma (k+2)}.  \label{8}
\end{equation}%
\noindent Then we can approximate Eq. (\ref{6}) by%
\begin{equation}
\tilde{E}_{int}(S_{ij})=\frac{R^{2}}{2LS_{ij}}\sum_{\lambda =1}^{\infty
}g_{\lambda }\text{ ,}  \label{9}
\end{equation}%
\noindent where $\lambda $ indicates the order of the expansion. As an
illustration, the first and second terms in the sum are%
\begin{equation}
g_{1}=1-\frac{1}{\alpha _{1}}\text{,}  \label{10-1}
\end{equation}%
and%
\begin{equation}
g_{2}=\frac{R^{2}}{4S_{ij}^{2}}\left( 1-\frac{\alpha _{2}}{\alpha _{1}}%
\right) +\frac{9R^{4}}{64S_{ij}^{4}}\left( 1-\frac{\alpha _{3}}{\alpha _{1}}%
\right) \text{,}
\end{equation}%
\noindent where $\alpha _{1}\equiv \sqrt{1+L^{2}/S_{ij}^{2}}$, $\alpha
_{2}\equiv (1-2L^{2}/S_{ij}^{2})/\alpha _{1}^{4}$ and $\alpha _{3}\equiv
(1+8L^{4}/3S_{ij}^{4}-8L^{2}/S_{ij}^{2})/\alpha _{1}^{8}$. Figure 1
illustrates the interaction energy between two identical nanowires with
parallel axial magnetization as a function of $2R/S_{ij}$. When the two
wires are in contact, $2R/S_{ij}=1$; when they are infinitely separated, $%
2R/S_{ij}=0$. In this figure the solid line represents the numerical
integration of the interaction energy, Eq. (\ref{6}), the dashed line is
given by the first order approximation of this energy, Eq. (\ref{9}) and the
dotted line corresponds to the second order approximation. From this figure
we observe that a first order approximation gives a reasonable approach to
Eq. (\ref{6}) for $L/R\gg 1$.\ We can conclude that the first term in the
expansion in Eq. (\ref{9}) gives a very good approach to Eq. (\ref{6}) for $%
2R/S_{ij}\leq 0.6$, and $L/R\gg 1$.

\begin{figure}[ht]
\begin{center}
\includegraphics[width=8cm,height=8cm]{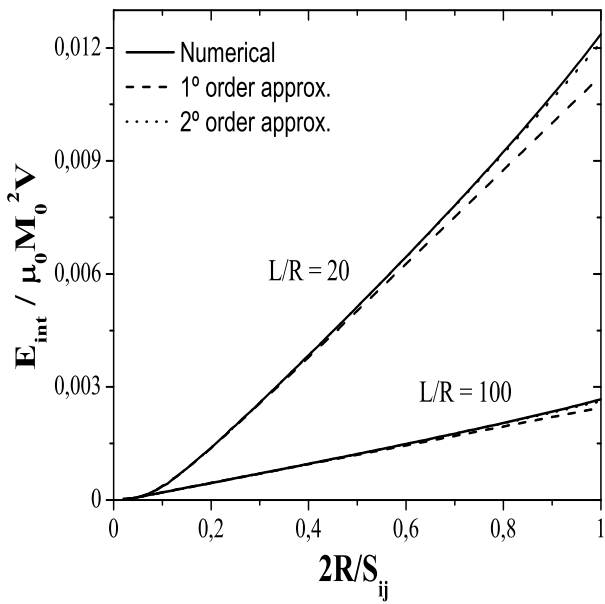}
\end{center}
\caption{Interaction energy between two identical nanowires with parallel
axial magnetization. The solid line corresponds to the numerical integration
of Eq. (5), the dashed line corresponds to the first order approximation of
Eq. (7) and the dotted line corresponds to the second order approximation of
Eq. (7).}
\label{f1}
\end{figure}

\noindent Also, for a large center-to-center distance between the wires, $%
S_{ij}\gg L$, we can expand $\alpha _{1}$\ in Eq. (\ref{10-1}), obtaining
the following expression for the interaction energy between two wires%
\begin{equation}
\tilde{E}_{\mu \text{-}\mu }(S_{ij})=\frac{V}{4\pi }\frac{1}{S_{ij}^{3}}%
\text{ ,}  \label{11}
\end{equation}%
\noindent This last expression, which we call the dipole-dipole
approximation, is equivalent to the interaction between two dipoles, that
is, each wires has been approximated by a single dipole.

In order to investigate the validity of this dipole-dipole ($\tilde{E}_{\mu
\text{-}\mu }$) approximation we calculate the ratio between the
magnetostatic interaction, Eq. (\ref{6}), and the dipole-dipole
approximation, Eq. (\ref{11}), between two identical nanowires as a function
of $\ 2R/S_{ij}$. These results are illustrated in Figure 2 and lead us to
conclude that the dipole-dipole approximation overestimates the real
interaction, except for very apart wires, which is not the usual case in an
array. Also for large aspect ratio wires the approximation becomes worst.

\begin{figure}[ht]
\begin{center}
\includegraphics[width=8cm,height=8cm]{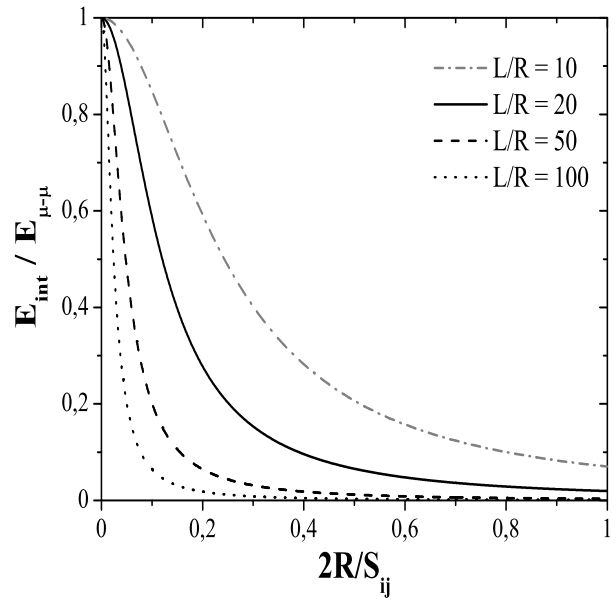}
\end{center}
\caption{Ratio between the magnetostatic interaction, Eq. (5), and the
dipole-dipole interaction, Eq. (10), as a function of $2R/S_{ij}$. Different
aspect ratio $L/R$ are considered.}
\label{FIG5}
\end{figure}

In order to quantify the importance of the interaction energy we calculated
the ratio between the self-energy and the magnetostatic interaction energy
between two identical nanowires,%
\begin{equation}
\eta =\frac{\tilde{E}_{int}(S_{ij})}{2\tilde{E}_{self}}\ .  \label{12}
\end{equation}%
\noindent Figure 3 defines the geometry of the two-wire-system for which $%
\eta =0.2,$ $0.1,$ $0.01$ and $0.001$. From this figure we observe a strong
dependence of the interaction energy on the geometry of the array. As an
illustration, when we consider two nanowires with $L=1$ $\mu $m, $R=20$ nm
and $L/R=50$, if we look for an almost non interacting regime, $\eta
=0.01,2R/S_{ij}=0.068,$ and then the two wires have to be at least $590$ nm
apart. For this geometry the interaction energy is about $1\%$\ of the self
energy. However, for the same $\ L$ and $R$, if the wires are $58$ nm apart (%
$2R/S_{ij}=0.69$),\ the interaction energy is about $20\%$ of the self
energy ($\eta =0.2)$.

\begin{figure}[ht]
\begin{center}
\includegraphics[width=8cm,height=8cm]{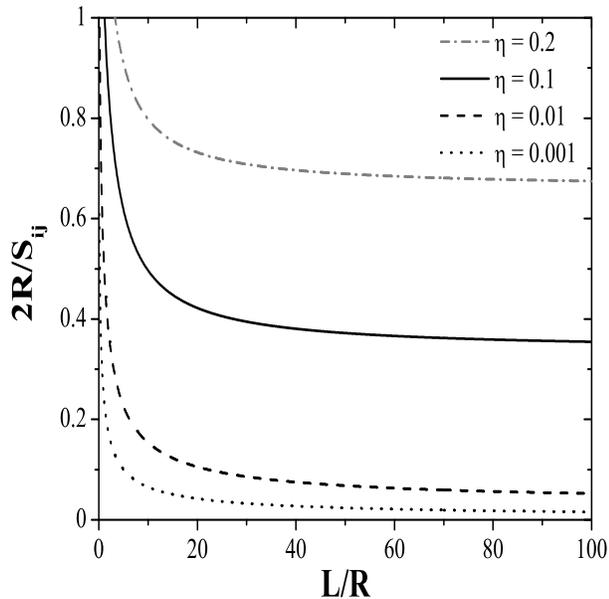}
\end{center}
\caption{$\protect\eta $ values as a function of the geometry of the
two-wire-system.}
\label{FIG6}
\end{figure}

\subsubsection{Wire array}

We are now in position to investigate the effect of the interwire
magnetostatic coupling in a square array. Calculations for the total
interaction energy $\tilde{E}_{int}^{\pm }(N)$ of the $N=n\times n$ square
array are shown in Appendix B, and lead us to write%
\begin{multline}
\tilde{E}_{array}^{\pm }(n)=2n\sum_{p=1}^{n-1}(n-p)(\pm 1)^{p}\tilde{E}%
_{int}(pd)  \label{12a} \\
+2\sum_{p=1}^{n-1}\sum_{q=1}^{n-1}(n-p)(n-q)(\pm 1)^{p-q}\tilde{E}_{int}(d%
\sqrt{p^{2}+q^{2}})\ ,
\end{multline}%
\noindent where $+$ ($-$) refers to parallel (antiparallel) magnetic
ordering of the nanowires in an array with nearest-neighbor distance $d$,
and $\tilde{E}_{int}$\ is the interaction energy between two wires given by
Eq. (\ref{6}). Note that in an array $S_{ij}$ is a function of \ $d$. In the
antiparallel configuration the magnetization of nearest-neighbor nanowires
points in opposite directions. Figure 4 illustrates the behavior of $%
W_{array}^{\pm }(n)=\tilde{E}_{array}^{\pm }(n)/n^{2}$ as a function of $n$
in a ferromagnetic (a) and an antiferromagnetic (b) array. We consider an
array of identical wires with $R=20$ nm and $L=1$ $\mu $m and two different
nearest-neighbor-distance $d$. We can see that a large number of wires ($%
N\approx 10^{6}$), corresponding to a sample of $\approx 0.01$ mm$^{2}$, is
required for reaching convergence of $W_{array}^{+}(n)$. However, in view of
cancellations originated in the different signs of the parallel and
antiparallel interactions, the antiparallel configuration converges faster,
requiring only the order of 10$^{2}$ wires and a sample of $\approx 1$ $\mu $%
m$^{2}$.

\begin{figure}[ht]
\begin{center}
\includegraphics[width=8cm,height=13cm]{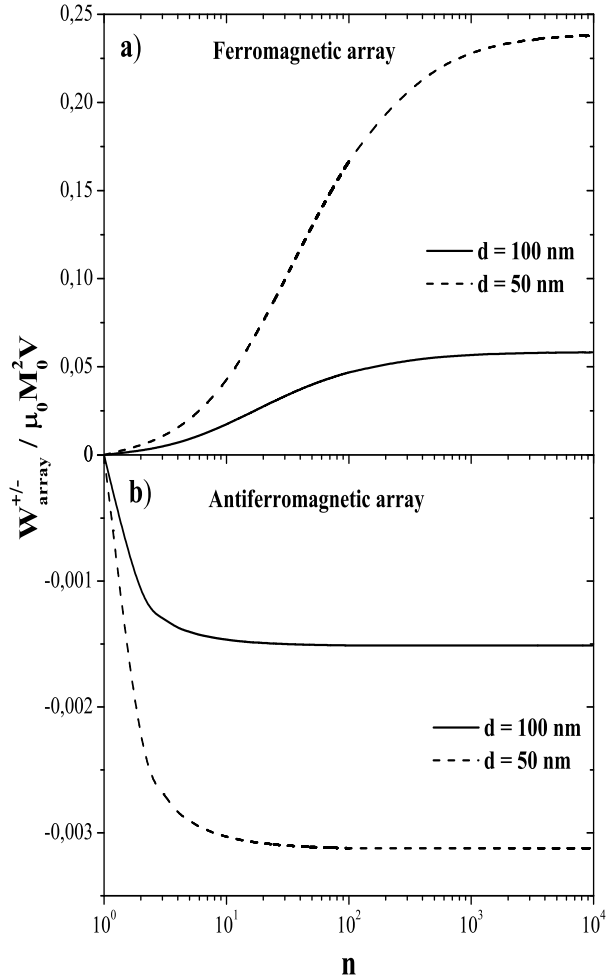}
\end{center}
\caption{$W_{array}^{\pm }(n)$ in a square array of identical wires ($R=20$
nm \ and $L=1$ $\protect\mu $m). We consider different number of wires in
the array and two nearest-neighbor-distances $d$.}
\label{FIG8}
\end{figure}

We also investigate, for the same wires ($R=20$ nm and $L=1$ $\mu $m), the
variation of the asymptotic value of $W_{array}^{\pm }(n)$ \ as a function
of the nearest-neighbor-distance $d$ in a ferromagnetic, (Fig. 5a), and
antiferromagnetic array, (Fig. 5b). Figure 5 illustrates our results showing
that in the ferromagnetic array, interaction effects decay in an exponential
way and extend over long distances as compared with $d.$ Figures 4 and 5
agree with conclusions from experiments by Nielsch \textit{et al}.\cite%
{Nielsch3} who assume that, due to the high aspect ratio of the magnetic
nanowires in an hexagonal array, the stray field interaction extends over
several nearest-neighbor-distances.

\begin{figure}[ht]
\begin{center}
\includegraphics[width=8cm,height=13cm]{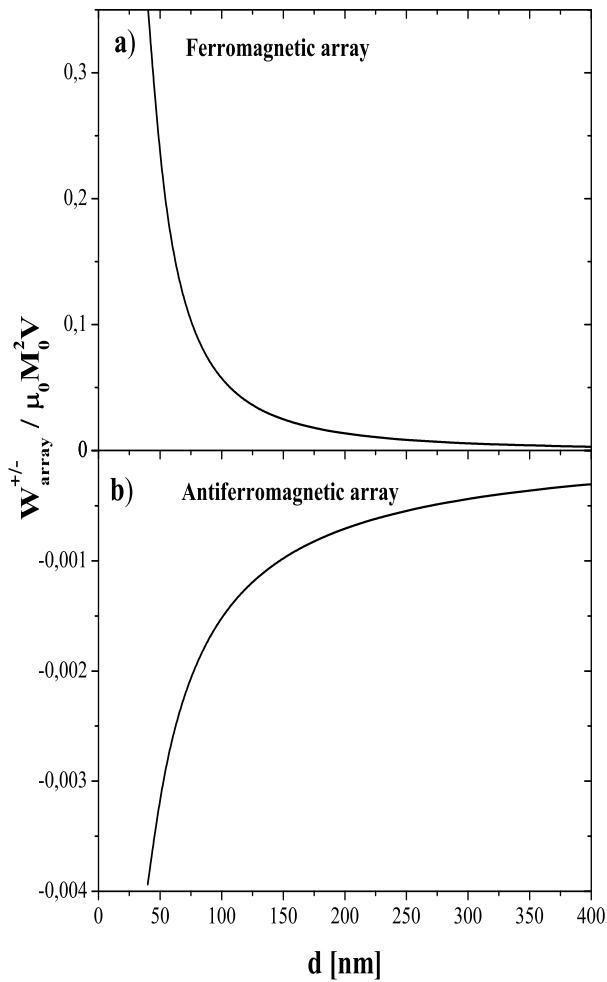}
\end{center}
\caption{Asymptotic values of the interaction energy in a square array
obtained with $N=10^{6}$ (ferromagnetic array) and $N=10^{2}$
(antiferromagnetic array). }
\label{FIG9}
\end{figure}

\noindent

\section{Concluding Remarks}

We have obtained analytical expressions for the magnetostatic interactions
between wires. By expanding these expressions we investigate first and
second order approximations to the interaction energies showing the range of
validity of these expansions. When the wires are apart distances much more
larger that their diameter, the first order approximation is valid. We also
conclude that the dipole-dipole interaction is valid only when $%
2R/S_{ij}\simeq 0$. The energy expressions lead us to investigate the
extension of the interwire magnetostatic interactions in the array. The
number of wires required to reach the convergence in the energy depends on
the relative magnetic ordering between nearest-neighbor wires. In the
parallel array, a big number of wires is required until convergence. This
result implies that the size of the array is an important factor to be
consider when different measurements have to be compared.

\section{Acknowledgments}

This work has been partially supported in Chile by FONDECYT 1050013 and
Millennium Science Nucleus \textquotedblleft Condensed Matter
Physics\textquotedblright\ P02-054F. CONICYT Ph.D. Fellowship and MECESUP
project USA0108 are also acknowledged. One of the authors, J.E., is grateful
to the Instituto de Ciencia de Materiales-CSIC for its hospitality.

\section{Appendix A}

In this appendix we calculate the interaction energy of two identical
nanowires. We start by replacing the functional form $\mathbf{M}_{i}(\mathbf{%
r})=M_{0}\sigma _{i}\mathbf{\hat{z}}$ in Eq. (\ref{2}), leading to\newline
\begin{equation}
E_{int}\left( i,j\right) =\mu _{0}\int\nolimits_{V}M_{iz}\frac{\partial U_{j}%
}{\partial z}dv\ .  \label{A1}
\end{equation}%
\noindent The magnetostatic potential is given by

\begin{equation*}
\mathbf{U}(\mathbf{r})=-\frac{1}{4\pi }\int\limits_{V}\frac{\nabla \cdot
\mathbf{M}(\mathbf{r}^{\prime })}{\left\vert \mathbf{r}-\mathbf{r}^{\prime
}\right\vert }\;dv^{\prime }+\frac{1}{4\pi }\int\limits_{S}\frac{\mathbf{%
\hat{n}}^{\prime }\cdot \mathbf{M}(\mathbf{r}^{\prime })}{\left\vert \mathbf{%
r}-\mathbf{r}^{\prime }\right\vert }\;ds^{\prime }\,,
\end{equation*}%
\noindent where $V$ and $S$ represent the volume and surface of the wire,
respectively.

Due to the functional form of $\mathbf{M}_{i}(\mathbf{r})$, the volumetric
contribution to the potential is zero. For the calculation of the surface
contribution to the magnetostatic potential, we use the following expansion
on the potential \cite{Jackson}%
\begin{equation*}
\frac{1}{\left\vert \mathbf{r}-\mathbf{r}^{\prime }\right\vert }%
=\sum_{m=-\infty }^{\infty }e^{im(\phi -\phi ^{\prime
})}\int\limits_{0}^{\infty }dkJ_{m}(kr_{i})J_{m}(kr^{\prime
})e^{-k(z_{>}-z_{<})}\ .
\end{equation*}%
\noindent\ Then, the surface contribution to the potential reads \cite%
{Landeros}%
\begin{multline*}
U_{j}=M_{0}\sigma _{j}\sum_{m=-\infty }^{\infty }\int\limits_{0}^{2\pi
}d\phi ^{\prime }e^{im(\phi -\phi ^{\prime })}\int\limits_{0}^{R}r^{\prime
}dr^{\prime } \\
\int\limits_{0}^{\infty }dkJ_{m}(kr_{i})J_{m}(kr^{\prime
})[e^{-k(L-z)}-e^{-kz}]\ \text{.}
\end{multline*}%
\noindent After some manipulations, it is straightforward to obtain the
potential at a distance $r_{i}$ from the axis of the wire $j$-$th$, $U_{j}$
\begin{equation*}
U_{j}=\frac{RM_{0}\sigma _{j}}{2}\int\limits_{0}^{\infty }\frac{dk}{k}%
J_{1}(kR)J_{0}(kr_{i})[e^{-k(L-z)}-e^{-kz}]\ .
\end{equation*}%
\noindent Now we need to calculate the dipolar field due to the wire $j$-$th$
experienced by the wire $i$-$th$ a distance $S_{ij}$ apart. In this case

\begin{equation*}
r_{i}=\sqrt{r^{2}+S_{ij}^{2}-2rS_{ij}\cos (\phi +\beta )}\ ,
\end{equation*}%
\noindent where $\beta $ is an arbitrary angle and $r$ defines a particular
point in wire $i$. Then, we include in the potential the following expansion
\cite{Jackson}%
\begin{multline*}
J_{0}\left( k\sqrt{r^{2}+S_{ij}^{2}-2rS_{ij}\cos (\phi +\beta )}\right) \\
=\sum_{m=-\infty }^{\infty }e^{im(\phi +\beta )}J_{m}(kr)J_{m}(kS_{ij})\ ,
\end{multline*}%
\noindent which lead us to obtain
\begin{multline*}
U_{j}=\frac{RM_{0}\sigma _{j}}{2}\int\limits_{0}^{\infty }\frac{dk}{k}%
J_{1}(kR)[e^{-k(L-z)}-e^{-kz}] \\
\sum_{m=-\infty }^{\infty }e^{im(\phi +\beta )}J_{m}(kr)J_{m}(kS_{ij})\ .
\end{multline*}%
\noindent By replacing this last expression in Eq. (\ref{A1}) we obtain%
\begin{multline*}
E_{int}(i,j)=2\mu _{0}M^{2}\sigma _{i}\sigma _{j}\pi R^{2} \\
\int\limits_{0}^{\infty }\frac{dk}{k^{2}}%
J_{1}^{2}(kR)J_{0}(kS_{ij})(1-e^{-kL})\ .
\end{multline*}%
\noindent Finally, and changing the variable $q=kL$, we obtain the reduced
expression presented in the Eq. (\ref{6}),\
\begin{equation*}
\tilde{E}_{int}(i,j)=2\sigma _{i}\sigma _{j}\int\limits_{0}^{\infty }\frac{dq%
}{q^{2}}J_{1}^{2}\left( \frac{qR}{L}\right) J_{0}\left( \frac{qS_{ij}}{L}%
\right) (1-e^{-q})\ .
\end{equation*}%
\ \ \ \ \ \ \ \ \ \ \ \ \ \ \ \ \

\section{Appendix B}

The total interaction energy of the $N=n\times n$ square array can be
written as
\begin{multline*}
\tilde{E}_{array}^{\pm }(n)=\frac{1}{2}\sum_{i=1}^{n}\sum_{j=1}^{n}%
\sum_{k=1}^{n}\sum_{l=1}^{n}(\pm 1)^{(k-i)-(l-j)} \\
\tilde{E}_{int}\left( d\sqrt{(k-i)^{2}+(l-j)^{2}}\right) \ ,
\end{multline*}%
\noindent where $+$ ($-$) refers to parallel (antiparallel) nearest-neighbor
magnetic orientation of the nanowires in the array. Here\ $\tilde{E}%
_{int}\left( 0\right) =0$, avoiding the self-interaction of the wires. For
simplicity we define the following function
\begin{equation}
f^{\pm }(p,q)=(\pm 1)^{p-q}\tilde{E}_{int}(d\sqrt{p^{2}+q^{2}})\ ,
\label{B2}
\end{equation}%
\noindent which can be used to write the interaction energy in a compact
form; that is%
\begin{equation}
\tilde{E}_{array}^{\pm }(n)=\frac{1}{2}\sum_{i=1}^{n}\sum_{j=1}^{n}%
\sum_{k=1}^{n}\sum_{l=1}^{n}f^{\pm }(k-i,l-j)\ .  \label{B3}
\end{equation}%
\noindent We can reduce the number of summations using the following rule%
\begin{equation}
\sum_{i=1}^{n}\sum_{k=1}^{n}g(k-i)=ng(0)+\sum_{p=1}^{n-1}(n-p)[g(p)+g(-p)]\ ,
\label{B4}
\end{equation}%
\noindent which lead us to write%
\begin{multline*}
\sum_{i=1}^{n}\sum_{k=1}^{n}f^{\pm }(k-i,l-j)=nf^{\pm }(0,l-j) \\
+2\sum_{p=1}^{n-1}(n-p)f^{\pm }(p,l-j)\ .
\end{multline*}%
\noindent Then, the interaction energy, Eq. (\ref{B3}), reduces to
\begin{multline}
\tilde{E}_{array}^{\pm }(n)=\frac{n}{2}\sum_{j=1}^{n}\sum_{l=1}^{n}f^{\pm
}(0,l-j)  \label{B5} \\
+\sum_{p=1}^{n-1}(n-p)\sum_{j=1}^{n}\sum_{l=1}^{n}f^{\pm }(p,l-j)\ .
\end{multline}%
\noindent Using again the rule (\ref{B4}), we can reduce the double-sums in
Eq. (\ref{B5}), obtaining%
\begin{multline*}
\tilde{E}_{int}^{\pm }(n)=\frac{n^{2}}{2}f^{\pm
}(0,0)+n\sum_{q=1}^{n-1}(n-q)f^{\pm }(0,q) \\
+n\sum_{p=1}^{n-1}(n-p)f^{\pm
}(p,0)+2\sum_{p=1}^{n-1}\sum_{q=1}^{n-1}(n-p)(n-q)f^{\pm }(p,q)\ .
\end{multline*}%
\noindent From (\ref{B2}) we know that $f^{\pm }(0,0)\equiv 0$, which lead
us to finally obtain
\begin{multline*}
\tilde{E}_{array}^{\pm }(n)=2n\sum_{p=1}^{n-1}(n-p)(\pm 1)^{p}\tilde{E}%
_{int}\left( pd\right) \\
+2\sum_{p=1}^{n-1}\sum_{q=1}^{n-1}(n-p)(n-q)(\pm 1)^{p-q}\tilde{E}%
_{int}\left( d\sqrt{p^{2}+q^{2}}\right) \ .
\end{multline*}

\end{document}